\documentclass[twocolumn, superscriptaddress, preprintnumbers, amsmath, amssymb]{revtex4}

\allowdisplaybreaks
\allowdisplaybreaks[4]

\usepackage{CJK}
\usepackage{graphicx}
\usepackage{dcolumn}
\usepackage{bm}
\usepackage[dvipdfm,
            pdfstartview=FitH,
            CJKbookmarks=true,
            bookmarksnumbered=true,
            bookmarksopen=true,
            colorlinks=true,
            pdfborder=001,
            linkcolor=blue,
            anchorcolor=blue,
            citecolor=blue
            ]{hyperref}

\begin{document}


\title{Static quadrupole moments of nuclear chiral doublet bands}

\author{Q. B. Chen}\email{qbchen@pku.edu.cn}
\affiliation{Physik-Department, Technische Universit\"{a}t
M\"{u}nchen, D-85747 Garching, Germany}

\author{N. Kaiser}\email{nkaiser@ph.tum.de}
\affiliation{Physik-Department,
             Technische Universit\"{a}t M\"{u}nchen,
             D-85747 Garching, Germany}

\author{Ulf-G. Mei{\ss}ner}\email{meissner@hiskp.uni-bonn.de}
\affiliation{Helmholtz-Institut f\"{u}r Strahlen- und Kernphysik and
             Bethe Center for Theoretical Physics, Universit\"{a}t Bonn,
             D-53115 Bonn, Germany}
\affiliation{Institute for Advanced Simulation,
             Institut f\"{u}r Kernphysik and J\"{u}lich Center for Hadron Physics,
             Forschungszentrum J\"{u}lich,
             D-52425 J\"{u}lich, Germany}
\affiliation{Ivane Javakhishvili Tbilisi State University, 0186
Tbilisi, Georgia}

\author{J. Meng}\email{mengj@pku.edu.cn}
\affiliation{State Key Laboratory of Nuclear Physics and Technology,
             School of Physics, Peking University,
             Beijing 100871, China}
\affiliation{Yukawa Institute for Theoretical Physics,
             Kyoto University,
             Kyoto 606-8502, Japan}

\date{\today}

\begin{abstract}
 The static quadrupole moments (SQMs) of nuclear chiral doublet bands are
 investigated for the first time taking the particle-hole configuration
 $\pi(1h_{11/2}) \otimes \nu(1h_{11/2})^{-1}$ with triaxial deformation
 parameters in the range $260^\circ \leq \gamma \leq 270^\circ$ as examples.
 The behavior of the SQM as a function of spin $I$ is illustrated by analyzing
 the components of the total angular momentum. It is found that in the
 region of chiral vibrations the SQMs of doublet bands are strongly varying
 with $I$, whereas in the region of static chirality the SQMs of doublet bands
 are almost constant. Hence, the measurement of SQMs provides a new criterion
 for distinguishing the modes of nuclear chirality. Moreover, in the high-spin
 region the SQMs can be approximated by an analytic formula with a proportionality
 to $\cos\gamma$ for both doublet bands. This provides a way to extract
 experimentally the triaxial deformation parameter $\gamma$ for chiral bands
 from the measured SQMs.
\end{abstract}

\maketitle


The phenomenon of nuclear chirality can appear in a fast
rotating nucleus with a triaxially deformed core and high-$j$
valence particles and holes~\cite{Frauendorf1997NPA}. In the
body-fixed frame, the angular momenta of the valence particles and
holes are aligned along the short and long axes of the triaxial
core, respectively, while the angular momentum of the rotational
core is aligned along the mediate axis. Then, the left-handed and
right-handed orientation of the three angular momenta are
degenerate, and a spontaneous breaking of this chiral symmetry may
happen. In the laboratory frame, due to the quantum mechanical
tunneling of the total angular momentum between the left-handed and
right-handed configurations, the chiral symmetry is, however, restored.
As a consequence, chiral doublet bands, i.e. pairs of nearly
degenerate $\Delta I=1$ bands with the same parity, are expected to
be observable~\cite{Frauendorf1997NPA}.

Up to now, more than 50 candidates for this phenomenon have been
observed in the mass regions $A\approx 80$, 100, 130, and 190. For
recent reviews on the subject, see Refs.~\cite{J.Meng2010JPG,
J.Meng2014IJMPE, Bark2014IJMPE, J.Meng2016PS, Raduta2016PPNP,
Starosta2017PS, Frauendorf2018PS, Q.B.Chen2020NPN} and the
corresponding data tables in Ref.~\cite{B.W.Xiong2019ADNDT}.
After the prediction~\cite{J.Meng2006PRC} and
confirmation~\cite{Ayangeakaa2013PRL} of multiple chiral doublet
bands in a single nucleus, the investigation of
chirality continues to be a hot topic in nuclear structure physics.

Besides the energy spectra, the electromagnetic transition strengths are
important observables for identifying nuclear chirality. Based on a model
with the configuration $\pi(1h_{11/2}) \otimes \nu(1h_{11/2})^{-1}$ and a
triaxial deformation parameter $\gamma=30^\circ$, the criteria for ideal nuclear
chirality are according to Refs.~\cite{Frauendorf1997NPA, Koike2004PRL,
Grodner2006PRL, Tonev2006PRL, Mukhopadhyay2007PRL, B.Qi2009PRC, J.Meng2010JPG,
Grodner2011PLB, Tonev2014PRL, Lieder2014PRL, Rather2014PRL, J.Meng2016PS}
similar intra-band and inter-band reduced magnetic dipole ($M1$) and
electric quadrupole ($E2$) transition strengths.

The search for additional observables that characterize nuclear chirality is
still an interesting question. Very recently, the first measurement of the
$g$-factor (gyromagnetic ratio) in a chiral band has been carried out for the
bandhead of $^{128}$Cs~\cite{Grodner2018PRL}. The $g$-factor can give
important information on the relative orientation of the three angular
momentum vectors of the particle, the hole, and the nuclear core. It
can also be used to discern whether the three angular momentum vectors lie
in a plane (planar configuration, known as chiral vibration) or
whether they span in the three-dimensional space (aplanar configuration,
known as static chirality).

In this work the static (electric) quadrupole moments (SQMs) [also
called spectroscopic quadrupole moments] of nuclear chiral doublet
bands will be investigated for the first time. As will be seen from
the following discussions, the SQM is related to the intrinsic
deformation parameter (a static property) and the orientation of the
total angular momentum (a dynamic property) of the nuclear system.
The SQM provides some essential information about the charge
distribution associated with the rotational motion, and it helps to
discern whether the angular momenta have formed configurations
related  to chiral vibration or static chirality.

Our calculations are based on the particle rotor model (PRM), which has
been widely used to describe chiral doublet bands and has achieved major
successes in this respect~\cite{Frauendorf1997NPA, J.Peng2003PRC, Koike2004PRL,
S.Q.Zhang2007PRC, B.Qi2009PLB, Lawrie2010PLB, Starosta2017PS, Q.B.Chen2018PRC,
Q.B.Chen2018PLB, Q.B.Chen2019PRC, Y.Y.Wang2019PLB}. It is a quantum mechanical model
that combines the collective rotational motion and the intrinsic single-particle
motions, describing the nuclear system in the laboratory frame. Its Hamiltonian
is diagonalized in states with the total angular momentum as a good quantum number.
The energy splitting and quantum mechanical tunneling probabilities between the
doublet bands can be obtained directly from the diagonalization process.
Actually, the basic input to the PRM can be obtained from covariant density
functional theory (CDFT)~\cite{J.Meng2006PRC, J.Meng2016book}, for practical
applications, see Refs.~\cite{Ayangeakaa2013PRL, Lieder2014PRL, Kuti2014PRL,
C.Liu2016PRL, Petrache2016PRC, Grodner2018PRL, Q.B.Chen2018PLB, J.Peng2019PLB,
B.F.Lv2019PRC}. Hence, the PRM can be used straightforwardly to investigate
the SQMs of chiral doublet bands.

In the PRM the Hamiltonian for a system with one proton and one neutron
coupled to a triaxial rigid (collective) rotor is composed
as~\cite{Frauendorf1997NPA, J.Peng2003PRC, Koike2004PRL, S.Q.Zhang2007PRC,
B.Qi2009PLB, Lawrie2010PLB, Starosta2017PS, Q.B.Chen2018PRC, Q.B.Chen2018PLB,
Q.B.Chen2019PRC, Y.Y.Wang2019PLB}
\begin{align}\label{eq1}
\hat{H}_\textrm{PRM}=\hat{H}_{\rm coll}+\hat{H}_{p}+\hat{H}_{n},
\end{align}
where $\hat{H}_{\rm coll}$ represents the Hamiltonian of the rigid rotor,
\begin{align}\label{eq6}
\hat{H}_{\rm coll}
  =\sum_{k=1}^3 \frac{\hat{R}_k^2}{2\mathcal{J}_k}
  =\sum_{k=1}^3 \frac{(\hat{I}_k-\hat{j}_{pk}-\hat{j}_{nk})^2}{2\mathcal{J}_k},
\end{align}
with the index $k=1, 2, 3$ referring to components along the three
principal axes in the body-fixed frame. Here, $\hat{R}_k$ and $\hat{I}_k$ are
the angular momentum operators of the collective rotor and
the total nucleus, while $\hat{j}_{p(n)k}$ is the angular momentum operator of
the valence proton (neutron). Moreover, the parameters $\mathcal{J}_k$ are the
three principal moments of inertia.

The Hamiltonians $\hat{H}_p$ and $\hat{H}_n$ describe a single
proton and neutron outside of the rotor. For a nucleon in a $j$-shell
orbital $\hat{H}_{p(n)}$ is given by
\begin{align}\label{eq2}
\hat{H}_{p(n)}=\pm \frac{C}{2}\Big\{\cos \gamma\Big[\hat{j}_3^2-\frac{j(j+1)}{3}\Big]
              +\frac{\sin \gamma}{2\sqrt{3}}\big(\hat{j}_+^2+\hat{j}_-^2\big)\Big\},
\end{align}
where the sign $\pm$ refers to a particle or hole
and $\gamma$ is the triaxial deformation parameter.
The coupling parameter $C$ is proportional to the quadrupole
deformation parameter $\beta$ of the rotor.

The PRM Hamiltonian in Eq.~(\ref{eq1}) can be solved by diagonalization in
the strong-coupling basis~\cite{Bohr1975, Ring1980book}
\begin{align}\label{eq3}
 & |j_p\Omega_p j_n\Omega_n K,IM\rangle \notag\\
 &=\sqrt{\frac{1}{2}}\Big[
 |j_p\Omega_p\rangle|j_n\Omega_n\rangle|IMK\rangle\notag\\
 &\quad +(-1)^{I-j_p-j_n}|j_p-\Omega_p\rangle|j_n-\Omega_n\rangle|IM-K\rangle\Big].
\end{align}
where $I$ denotes the total angular momentum quantum number of
the odd-odd nuclear system (rotor plus proton and neutron) and $M$ ($K$) refers
to the projection onto the $z$-axis (3-axis) in the laboratory (intrinsic) frame.
Furthermore, $\Omega_{p(n)}$ is the quantum number for the 3-axis component of the
valence nucleon angular momentum operator $\bm{j}_{p(n)}$ in the intrinsic frame,
while the states $|IMK\rangle$ are represented in terms of three Euler angles
$(\psi^\prime, \theta^\prime,\varphi^\prime)$ by the conventional Wigner-functions
$\sqrt{\frac{2I+1}{8\pi^2}}D_{M,K}^I(\psi^\prime, \theta^\prime,\varphi^\prime)$.
Under the requirement of the $\textrm{D}_2$ symmetry of a triaxial
nucleus~\cite{Bohr1975}, $K$ and $\Omega_{p}$ take the values: $K=-I,\dots,I$
and $\Omega_p=-j_p, \dots, j_p$. The quantum number $\Omega_n$ goes over the range
$\Omega_n=-j_n, \dots, j_n$ and it has to fulfil the condition that $K_R=K-\Omega_p-\Omega_n$ is
a positive even integer.

The PRM eigenfunctions are expressed in the strong-coupling basis as
\begin{align}
 |IM\rangle
 &=\sum_{K\Omega_p\Omega_n}f_{IK\Omega_p\Omega_n} |j_p\Omega_p j_n\Omega_n K,IM\rangle,
\end{align}
where the coefficients $f_{IK\Omega_p\Omega_n}$ are obtained by diagonalizing
the Hamiltonian $\hat{H}_\textrm{PRM}$. With the obtained wave functions
the SQMs are calculated as~\cite{Bohr1975, Ring1980book}
\begin{align}
 Q(I)=\langle II|\hat{Q}_{20}|II\rangle,
\end{align}
where the quadrupole momentum operator in the laboratory frame $\hat{Q}_{20}$
is related to the intrinsic quadrupole moment $Q_{2\nu}^\prime$ by
\begin{align}\label{eq7}
 \hat{Q}_{20}=\sum_{\nu} D_{0,\nu}^{2} Q_{2\nu}^\prime,
\end{align}
with $Q_{20}^\prime =Q_0^\prime\cos\gamma$, $Q_{21}^\prime =Q_{2-1}^\prime=0$,
$Q_{22}^\prime =Q_{2-2}^\prime=Q_0^\prime\sin\gamma/\sqrt{2}$.
Here, $Q_0^\prime$ is an empirical quadrupole moment that is related
to the axial deformation $\beta$ by  $Q_0^\prime=3R_0^2Z\beta/\sqrt{5\pi}$,
where $Z$ is the proton number and $R_0=1.2\,{\rm fm}\,A^{1/3}$.
One can finally obtain the SQM for each band as
\begin{align}
 Q(I)
 &=\langle II20|II\rangle
 \sum_{\Omega_p\Omega_n}\sum_{KK^\prime}
 f_{IK\Omega_p\Omega_n}^* f_{IK^\prime \Omega_p\Omega_n}\notag\\
 &\quad \times \sum_\nu \langle IK^\prime2\nu|IK\rangle Q_{2\nu}^\prime.
\end{align}

The computation of the SQM is straightforward with the given PRM wave function.
In the following, we give two alternative ways of calculating the SQM.

On the one hand, one notices that the relevant Wigner-functions $D_{0,\nu}^2$ in
Eq.~(\ref{eq7}) serve as eigenvalues of certain angular momentum
operators when acting on the states $|II\rangle$ \cite{Marshalek1977NPA}
\begin{align}
 D_{0,0}^2 |II\rangle &=\frac{3\hat{I}_3^2-I(I+1)}{(I+1)(2I+3)}|II\rangle,\\
 D_{0,2}^2|II\rangle  &=\sqrt{\frac{3}{2}}\frac{\hat{I}_+^2}{(I+1)(2I+3)}|II\rangle,\\
 D_{0,-2}^2 |II\rangle&=\sqrt{\frac{3}{2}}\frac{\hat{I}_-^2}{(I+1)(2I+3)}|II\rangle,
\end{align}
with the raising and lowering operators $\hat{I}_{\pm}=\hat{I}_1 \pm i\hat{I}_2$.
Therefore, one gets a decomposition into two contributions
\begin{align}\label{eq5}
 Q(I) &=Q_0(I)+Q_2(I),\\
 \label{eq10}
 Q_0(I)&=\frac{3\langle \hat{I}_3^2\rangle-I(I+1)}{(I+1)(2I+3)}Q_0^\prime \cos\gamma,\\
 \label{eq11}
 Q_2(I)&=\frac{\sqrt{3}(\langle \hat{I}_1^2\rangle-\langle \hat{I}_2^2\rangle)}
        {(I+1)(2I+3)}Q_0^\prime\sin\gamma.
\end{align}
In the case of prolate deformation $\gamma=0^\circ$, and if the third component
of the angular momentum $\langle \hat{I}_3\rangle=K$ is a good quantum number,
the part $Q_2(I)$ vanishes and $Q(I)$ becomes
\begin{align}
  Q(I)&=\frac{3K^2-I(I+1)}{(I+1)(2I+3)}Q_0^\prime.
\end{align}
This famous formula has already been given in textbooks (e.g.,
Refs.~\cite{Bohr1975, Ring1980book}). It is often used to extract
the intrinsic quadrupole moment $Q_0^\prime$ of an individual
state from the measured $Q(I)$-values for an axially deformed nucleus.

On the other hand, if the $z$-axis in the laboratory frame is chosen along the
angular momentum $\bm{I}$ (which is realized by $M=I$), the Euler angles
$(\psi^\prime, \theta^\prime,\varphi^\prime)$ and the tilted angles
$(\theta, \varphi)$ are related in the following way
\begin{align}
 \theta=\theta^\prime, \quad \varphi=\pi-\varphi^\prime.
\end{align}
Here, $\theta$ is the angle between the total spin $\bm{I}$ and the
3-axis, and the $\varphi$ is the angle between the projection of total
spin $\bm{I}$ onto the 12-plane and the 1-axis. With this connection
one can express the quadrupole moment operator $\hat{Q}_{20}$ a function
of $\theta$ and $\varphi$.

The Wigner-functions at $\psi'=0$ have the form
\begin{align}
 D_{0,0}^2(0,\theta^\prime,\varphi^\prime)&=\frac{1}{2}(3\cos^2\theta-1),\\
 D_{0,2}^2(0,\theta^\prime,\varphi^\prime)&=\sqrt{\frac{3}{8}}\sin^2\theta e^{-2i\varphi},\\
 D_{0,-2}^2(0,\theta^\prime,\varphi^\prime)&=\sqrt{\frac{3}{8}}\sin^2\theta e^{2i\varphi},
\end{align}
and hence,
\begin{align}\label{eq4}
 \hat{Q}_{20}(\theta,\varphi)
&= \frac{1}{2}(3\cos^2\theta-1)Q_0^\prime\cos\gamma \notag\\
&\quad+ \frac{\sqrt{3}}{2}\sin^2\theta (\cos^2\varphi-\sin^2\varphi)Q_0^\prime\sin\gamma.
\end{align}
Once the probability distribution of the orientation of the total
angular momentum $\mathcal{P}(\theta,\varphi)$ (called azimuthal
plot~\cite{F.Q.Chen2017PRC, Q.B.Chen2018PRC_v1, Streck2018PRC})
is known, the SQM can be calculated as a solid angle integral
\begin{align}
 Q(I)=\int_0^\pi \sin\theta d\theta \int_{0}^{2\pi} d\varphi~\hat{Q}_{20}(\theta,\varphi)
 \mathcal{P}(\theta,\varphi).
\end{align}

Combining Eqs.~(\ref{eq5}) and (\ref{eq4}), one can find interesting
relationships between expectation values of the tilted angles
$(\theta,\varphi)$ and the angular momentum components $\hat{I}_k$,
\begin{align}
\label{eq12}
 \langle \sin^2\theta \cos^2\varphi \rangle
 &=\frac{\langle \hat{I}_1^2\rangle+(I+1)/2}{(I+1)(I+3/2)},\\
\label{eq13}
 \langle \sin^2\theta \sin^2\varphi \rangle
 &=\frac{\langle \hat{I}_2^2\rangle+(I+1)/2}{(I+1)(I+3/2)},\\
\label{eq14}
  \langle \cos^2\theta \rangle
  &=\frac{\langle \hat{I}_3^2\rangle+(I+1)/2}{(I+1)(I+3/2)}.
\end{align}
It follows immediately that
\begin{align}
\langle \sin^2\theta
\cos^2\varphi \rangle+\langle \sin^2\theta \sin^2\varphi \rangle
+ \langle \cos^2\theta \rangle=1
\end{align}
holds, consistent with the normalization of the PRM wave function.
Especially, at $I=0$ one has $\langle \hat{I}_k^2\rangle=0$ and gets an
isotropic angular distribution $\langle \sin^2\theta \cos^2\varphi \rangle=\langle
\sin^2\theta \sin^2\varphi \rangle
=\langle \cos^2\theta \rangle=1/3$.

\begin{figure}[!ht]
  \begin{center}
    \includegraphics[width=7.0 cm]{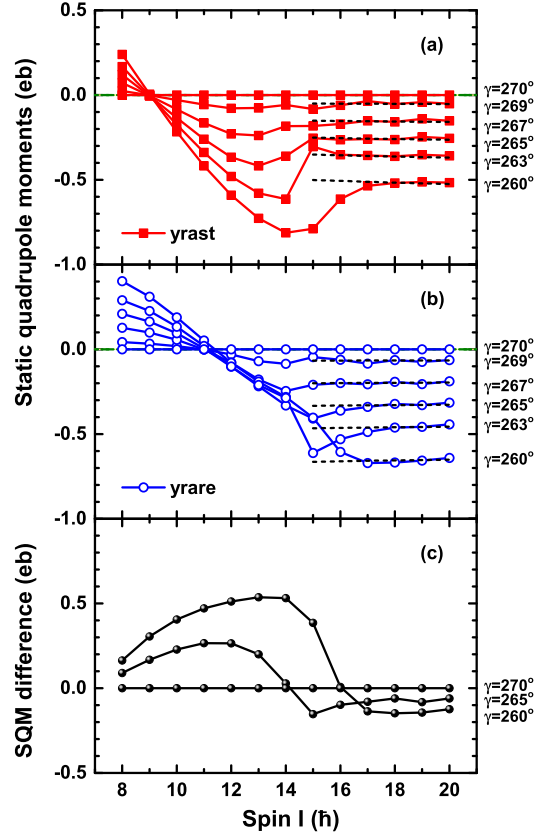}
    \caption{Static quadrupole moments of yrast (a) and yrare (b) bands
    calculated in the PRM for the configuration $\pi(1h_{11/2})^1\otimes
    \nu(1h_{11/2})^{-1}$ with $\gamma=270^\circ$, $269^\circ$, $267^\circ$,
    $265^\circ$, $263^\circ$, and $260^\circ$. The lines for the yrast band
    are given by Eq.~(\ref{eq8}) and for the yrare band by Eq.~(\ref{eq9}).
    (c) The corresponding  difference of static quadrupole moments between the
    yrare and yrast bands at $\gamma=270^\circ$, $265^\circ$, $260^\circ$.}\label{fig1}
  \end{center}
\end{figure}

In the calculations of the chiral doublet bands for the configuration
$\pi(1h_{11/2})^1\otimes\nu(1h_{11/2})^{-1}$ the deformation parameter $\beta=
0.23$ is fixed and the chosen coupling coefficients are $C_{p}=0.32$~MeV
and $C_n=-0.32$~MeV, in accordance with standard values for the mass region
$A\approx 130$. For the moments of inertia, the irrotational flow formula
$\mathcal{J}_k=\mathcal{J}_0\sin^2(\gamma-2k\pi/3)$ with
$\mathcal{J}_0=30~\hbar^2/\textrm{MeV}$ is used. In addition, the value
$Q_0^\prime=3.5~e\textrm{b}$ is chosen for the empirical electric quadrupole
moment.

The present studies focus on following values of the triaxial deformation
parameter: $\gamma=270^\circ$, $269^\circ$, $267^\circ$, $265^\circ$, $263^\circ$,
and $260^\circ$. With these specifications of $\gamma$, the 1-axis, 2-axis, and
3-axis are the short ($s$), long ($l$), and mediate ($m$) axes
of the triaxially deformed ellipsoid, respectively. Note that the moment of
inertia  $\mathcal{J}_m$ with respect to the $m$-axis is the largest for all
selected $\gamma$-values. The moment of inertia $\mathcal{J}_s$ is equal to
$\mathcal{J}_l$ at $\gamma=270^\circ$, while it is a bit larger than
$\mathcal{J}_l$ for the other $\gamma$-values. Moreover, for
the particle-hole configuration $\pi(1h_{11/2})^1\otimes\nu(1h_{11/2})^{-1}$
with this range of $\gamma$, there occurs the so-called chiral geometry
in a certain spin-region, according to the investigations
presented in Refs.~\cite{Frauendorf1997NPA, B.Qi2009PRC, Q.B.Chen2010PRC,
H.Zhang2016CPC, Q.B.Chen2018PRC, Q.B.Chen2018PRC_v1, Q.B.Chen2019PRC}.
We refer to these papers for the corresponding results concerning
energy spectra, electromagnetic transition probabilities,
and the entire angular momentum geometry.

\begin{figure}[!ht]
  \begin{center}
    \includegraphics[width=7.0 cm]{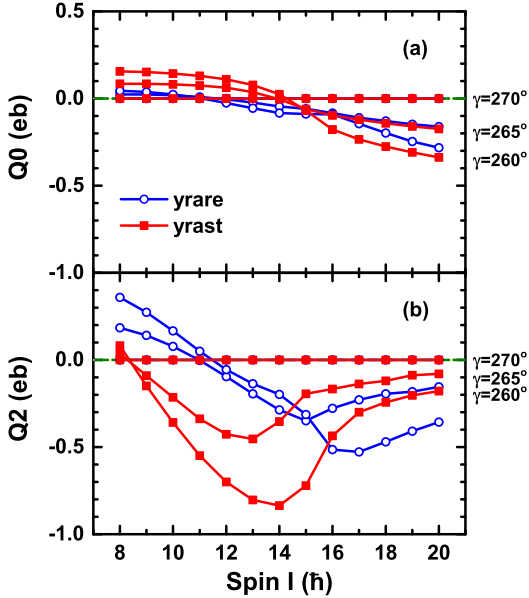}
 \caption{Contributions to the static quadrupole moment $Q_0(I)$ (a) and $Q_2(I)$
 (b) for doublet bands calculated in the PRM for the particle-hole
 configuration $\pi(1h_{11/2})^1\otimes \nu(1h_{11/2})^{-1}$ with $\gamma=
 270^\circ$, $265^\circ$, and $260^\circ$.}\label{fig3}
  \end{center}
\end{figure}

The corresponding SQMs of yrast and yrare bands as calculated in the PRM
are shown in Fig.~\ref{fig1}. Its contributions $Q_0(I)$ and $Q_2(I)$ as
calculated by Eqs.~(\ref{eq10}) and (\ref{eq11}) are shown in Fig.~\ref{fig3}.

\begin{figure}[!ht]
  \begin{center}
    \includegraphics[width=7.0 cm]{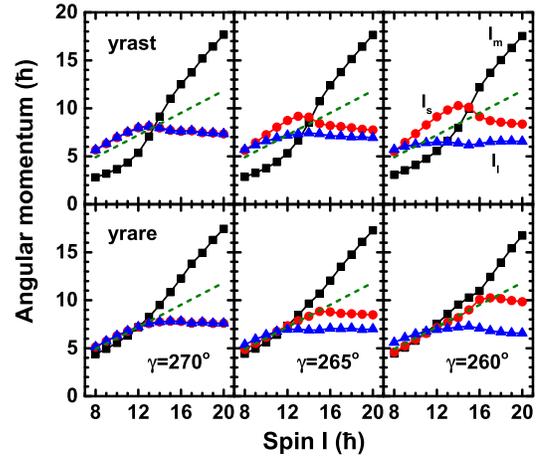}
    \caption{Root mean-square values of the total angular
    momentum components along the short ($s$-, circles),
    long ($l$-, triangles), and intermediate ($m$-, squares) axes as functions
    of the spin $I$ in the PRM for doublet bands at $\gamma=270^\circ$,
    $265^\circ$, and $260^\circ$. The dashed lines represent the average quantity
    $\sqrt{I(I+1)/3}$.}\label{fig2}
  \end{center}
\end{figure}

These two figures display the variation of the SQMs with the triaxial
deformation parameter $\gamma$. At $\gamma=270^\circ$, the SQMs of the yrast
and yrare bands are both zero over the entire spin region. The vanishing
values of $Q_0(I)$ result from the fact that $Q_0(I)$ and $Q_2(I)$ are both
zero. Note that $Q_0(I)=0$ stems from the static property $\cos 270^\circ=0$,
and $Q_2(I)=0$ has a dynamical origin, namely $\langle \hat{I}_s^2\rangle=
\langle\hat{I}_l^2\rangle$ as will be demonstrated in Fig.~\ref{fig2}.

When $\gamma$ deviates from $270^\circ$, the values of $Q(I)$ do not
vanish any longer. At several low-spins, the SQMs come out positive.
With increasing spin, $Q(I)$ decreases first, then shows a rapid increase,
and finally becomes almost constant. One can observe that with decreasing
deformation parameter $\gamma$, the static quadrupole moment $Q(I)$
decreases in the high-spin region.

An interesting finding about $Q(I)$ at high-spin $I$ is that it can be
well approximated by an analytic formula for both doublet bands.
For the yrast band, the formula reads
\begin{align}\label{eq8}
 Q(I)=\frac{3I^2-I(I+1)}{(I+1)(2I+3)}Q_0^\prime\cos\gamma,
\end{align}
while for the yrare band, it is
\begin{align}\label{eq9}
 Q(I)=\frac{3(I+3/2)^2-I(I+1)}{(I+1)(2I+3)}Q_0^\prime\cos\gamma.
\end{align}
Using these relations, one can extract the triaxial deformation parameter
(located in the range $240^\circ\leq \gamma \leq 300^\circ$)
from the experimentally measured $Q(I)$-values as
\begin{align}
\gamma=\arccos \bigg\{\frac{Q(I)}{Q_0^\prime}\frac{(I+1)(2I+3)}{3I^2-I(I+1)}\bigg\}
\end{align}
for the yrast band, and
\begin{align}
 \gamma=\arccos \bigg\{ \frac{Q(I)}{Q_0^\prime}\frac{(I+1)(2I+3)}{3(I+3/2)^2-I(I+1)} \bigg\}
\end{align}
for the yrare band, assuming a common value of $Q_0^\prime$.

Moreover, the above connections together with Eqs.~(\ref{eq5})-(\ref{eq11})
suggest that the three expectation values $\langle \hat{I}_k^2\rangle$ of
(squared) angular momentum components satisfy in the high-spin region the
(approximate) relations,
\begin{align}
 \frac{\sqrt{3}(I^2-\langle\hat{I}_m^2\rangle)}
 {\langle \hat{I}_s^2\rangle-\langle \hat{I}_l^2\rangle} \approx \tan\gamma
\end{align}
for the yrast band, and
\begin{align}
 \frac{\sqrt{3}[(I+3/2)^2-\langle\hat{I}_m^2\rangle]}
 {\langle \hat{I}_s^2\rangle-\langle \hat{I}_l^2\rangle} \approx \tan\gamma
\end{align}
for the yrare band.

In order to understand better the behavior of the $Q(I)$-values in
Fig.~\ref{fig3}, we have also shown the individual contributions
$Q_0(I)$ and $Q_2(I)$ at $\gamma=270^\circ$, $265^\circ$, and $260^\circ$.
One can see that generally $Q_0(I)$ is much smaller than $Q_2(I)$,
due to the suppression by the factor $\cos\gamma$ (a static property).
Hence, the behavior of $Q(I)$ as a function of spin $I$
is mainly determined by $Q_2(I)$. In particular, this explains
the decreasing trend visible in the low-spin region.

\begin{figure}[!ht]
  \begin{center}
    \includegraphics[width=7.0 cm]{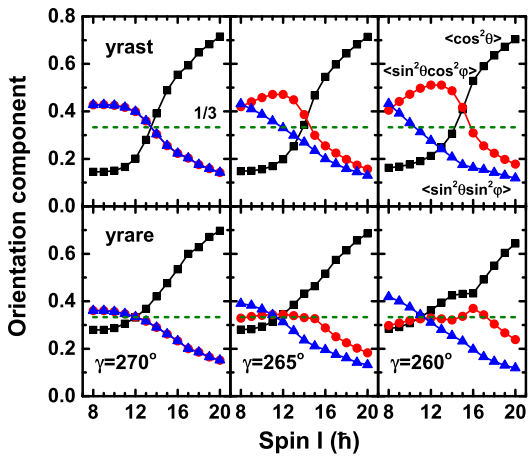}
    \caption{Expectation values of
    $\langle \sin^2\theta\cos^2\varphi\rangle$,
    $\langle \sin^2\theta\sin^2\varphi\rangle$, and
    $\langle \cos^2\theta\rangle$ derived from Eqs.~(\ref{eq12})-(\ref{eq14})
    for the doublet bands at $\gamma=270^\circ$, $265^\circ$, and $260^\circ$.
    The dashed line represents the isotropic average value $1/3$.}\label{fig4}
  \end{center}
\end{figure}

In order to present more details, we show in Fig.~\ref{fig2} the root
mean-square values of the total angular momentum component along the $s$-axis,
$l$-axis, and $m$-axis ($I_s=\langle \hat{I}_s^2\rangle^{1/2}$, $I_l=\langle
\hat{I}_l^2\rangle^{1/2}$, and $I_m=\langle \hat{I}_m^2\rangle^{1/2}$) as
functions of the spin $I$ for the doublet bands at $\gamma=270^\circ$,
$265^\circ$, and $260^\circ$. The corresponding orientation components, defined
as the expectation values $\langle \sin^2\theta\cos^2\varphi\rangle$,
$\langle \sin^2\theta\sin^2\varphi\rangle$, and $\langle \cos^2\theta\rangle$
according to Eqs.~(\ref{eq12})-(\ref{eq14}) are displayed in Fig.~\ref{fig4}.
The dashed lines at height $\sqrt{I(I+1)/3}$ and $1/3$ refer to completely
isotropic distributions.

Obviously, the equalities $I_s=I_l$ and $\langle \sin^2\theta\cos^2\varphi
\rangle = \langle \sin^2\theta \sin^2\varphi\rangle$ hold at $\gamma=270^\circ$.
This is a consequence of the symmetric configuration of the proton-particle
(mainly aligning along the $s$-axis) and the neutron-hole (mainly aligning
along the $l$-axis) as well as the equivalence of the moments of inertia with
respect to the $s$-axis and $l$-axis. As mentioned above, this coincidence
causes $Q_2(I)=0$ at $\gamma=270^\circ$.

When $\gamma$ deviates from $270^\circ$, $I_s$ is no longer equal to
$I_l$. For the yrast band, $I_s$ is in general larger than $I_l$
since $\mathcal{J}_s>\mathcal{J}_l$. Consequently, $Q_2(I)$ of
the yrast band becomes negative, noting that $\sin\gamma$ is negative.
For the yrare band, $I_s$ is smaller than $I_l$ in the region
$I\leq 12\hbar$. Correspondingly, the contribution $Q_2(I)$ as well as the total
$Q(I)$ are positive as shown in Figs.~\ref{fig3} and \ref{fig1}. At
$I\geq 13\hbar$, one has $I_s>I_l$, similar to the situation in the yrast band.
These features generate the (chiral) picture of a left-handed and a right-handed
configuration in the body-fixed frame in the spin region $15 \leq I\leq
18\hbar$.

The component $I_m$ increases with spin. For the yrast band, in the region
with $I\leq 14\hbar$ and $I_m<\sqrt{I(I+1)/3}$, corresponding to $\langle
\cos^2\theta\rangle<1/3$, the total angular momentum is located
close to the $sl$-plane. This gives positive the $Q_0(I)$-values as
shown in Fig.~\ref{fig3} (noting that $\cos\gamma$ is negative). In
this region, $I_m$ is smaller than $I_s$ and $I_l$. This is
due that the collective rotor motion has just started at the
bandhead of the doublet bands. The component $I_m$ is larger in the yrare
band than in the yrast band, which can be  attributed to the oscillations
of the total angular momentum between the left-handed and the right-handed
configuration (about the $sl$-plane). This phenomenon is known as the chiral
vibration. Accordingly, the contribution $Q_0(I)$ of the static quadrupole moment
is smaller in the yrare band than in the yrast band,
as shown in Fig.~\ref{fig3}. In the spin region $15 \leq I\leq 18\hbar$,
the component $I_m$ and $\langle \cos^2\theta \rangle$ behave similarly in
both doublet bands, which corresponds to the phenomen of static chirality.
Consequently, the static quadrupole moments $Q(I)$ in both doublet bands are
close to each other in this spin region.

In Fig.~\ref{fig1}, we show furthermore the difference of SQMs between
the yrare and yrast bands at deformation parameters $\gamma=270^\circ$,
$265^\circ$, and $260^\circ$. According to the above analysis, when
$\gamma$ deviates from $270^\circ$, the differences of the SQMs between
the doublet bands can be interpreted as the chiral vibration, and the
similar SQMs are attributed to the static chirality. Therefore, a
measurement of static quadrupole moments can provide a new criterion
for nuclear chirality.

In summary, the SQMs of chiral doublet bands have been investigated for the
first time taking the particle-hole configuration $\pi(1h_{11/2}) \otimes
\nu(1h_{11/2})^{-1}$ with triaxial deformation parameters in the range
$260^\circ \leq \gamma \leq 270^\circ$ as examples.
The behavior of the SQMs as a function of spin $I$ is illustrated
by analyzing  the components of the total angular momentum. Pronounced
differences of the SQMs between the doublet bands are attributed to
the chiral vibration, whereas their similarity signifies the static chirality.
This provides a new criterion to distinguish the modes of nuclear chirality.
Moreover, it is found that in the high-spin region the SQMs can be approximated
by an analytic formula with a proportionality to $\cos\gamma$ for both
doublet bands. It provides a way to extract experimentally the triaxial
deformation parameter $\gamma$ of chiral bands from the measured SQMs.
In view of this connection, experimental measurements of SQMs
for the chiral doublet bands are strongly suggested.

\section*{Acknowledgements}

One of the authors (Q.B.C.) thanks S. Frauendorf and I. Hamamoto for helpful
discussions. This work has been supported in parts
by Deutsche Forschungsgemeinschaft (DFG) and National Natural Science
Foundation of China (NSFC) through funds provided by the Sino-German
CRC 110 ``Symmetries and the Emergence of Structure in QCD''
(DFG Grant No. TRR110 and NSFC Grant No. 11621131001), the
National Key R\&D Program of China (Contract No. 2017YFE0116700 and
No. 2018YFA0404400), the NSFC under Grant No. 11935003, and
the State Key Laboratory of Nuclear Physics and Technology of Peking
University (Grant No. NPT2020ZZ01). The work of U.-G.M.
was also supported by the Chinese Academy of Sciences (CAS) through a
President's International Fellowship Initiative (PIFI) (Grant No. 2018DM0034)
and by the VolkswagenStiftung (Grant No. 93562).



\end{document}